# Optical Metrology of Sub-Wavelength Objects Enabled by Artificial Intelligence

Carolina Rendón-Barraza[1], Eng Aik Chan[1], Guanghui Yuan[1], Giorgio Adamo[1], Tanchao Pu[2] and Nikolay I. Zheludev[1,2,*]

[1]Centre for Disruptive Photonic Technologies, The Photonics Institute, School of Physical and Mathematical Sciences, Nanyang Technological University, 637371 Singapore.

[2]Centre for Photonic Metamaterials and Optoelectronics Research Centre, University of Southampton, Southampton SO17 1BJ, United Kingdom.

*Correspondence to: zheludev@soton.ac.uk.

**Abstract.** Microscopes and various forms of interferometers have been used for decades in optical metrology of objects that are typically larger than the wavelength of light λ. However, metrology of subwavelength objects was deemed impossible due to the diffraction limit. We report that measurement of the physical size of sub-wavelength objects with accuracy exceeding λ/800 by analyzing the diffraction pattern of coherent light scattered by the objects with deep learning enabled analysis. With a 633nm laser, we show that the width of sub-wavelength slits in opaque screen can be measured with accuracy of 0.77nm, challenging the accuracy of electron beam and ion beam lithographies. The technique is suitable for high-rate non-contact measurements of nanometric sizes in smart manufacturing applications with integrated metrology and processing tools.

Accurate measurements of a subwavelength object under conventional microscope by analyzing its image are impossible because the image blurs. Advanced nonlinear and statistical optical techniques such as the stimulated emission depletion (STED) and single-molecule localization methods (SMLM) can measure sub-wavelength objects (*1, 2*) but are generally unsuitable for non-invasive metrology, as they require impregnation of luminescent molecules or quantum dots into the sample. Currently, only imaging techniques using focused beams of particles with small De Broglie wavelength such as accelerated electrons in electron microscopes and ions in ion microscopes can be used to measure nanoscale dimensions. However, they are inherently slow due to the scanning nature of imaging and are unsuitable for smart manufacturing applications as they require samples in vacuum. Mutual displacements of macroscopic objects on the nanoscale can be measured using a recently developed "optical ruler" technology based on interferometry of light's singularities (*3*). Furthermore, it has been demonstrated that dimension-retrieval of subwavelength objects is possible by mapping the intensity profile of the diffraction pattern of scattered light from the object using an artificial neural network trained on data of similar objects with *a priori* known dimensions (*4*). Random dimers of two subwavelength slits have been measured with accuracy of *λ/10*. Here we report nearly two orders of magnitude improvement in measurement demonstrating accuracy exceeding *λ/800*.

We measured slits of random width that are cut in an opaque screen. Each slit was placed at a random position along *x*-direction within a rectangular frame, defined by four alignment marks. Each slit is characterized by its width *W* and displacement *D* from the center of its rectangular frame (Fig. 1).

Our experiments were performed in a dual-mode optical microscope (Fig.1). The sample with the slits was placed at the imaging plane of the apparatus and illuminated through a low numerical aperture lens (NA = 0.1) with a coherent light source at wavelength $\lambda$ = 633nm. Light diffracted from the sub-wavelength slit was then imaged (mapped) by a high-numerical aperture lens (NA = 0.9) at distances of $H = 2\lambda$, $5\lambda$ and $10\lambda$ from the screen and at the screen level ($H = 0$). An imaging system with a 4X magnification changer and a 5.5-megapixel sCMOS camera with 6.5μm pixel size was used. Since the diffracted field reaching the image sensor is formed by free-space propagating waves, it can be imaged at any magnification without loss of resolution by adjusting the magnification level necessary to ensure that the detector pixels are smaller than the required resolution. Our imaging system had a magnification of 333X corresponding to an effective pixel size of 19.5nm on the reference plane.

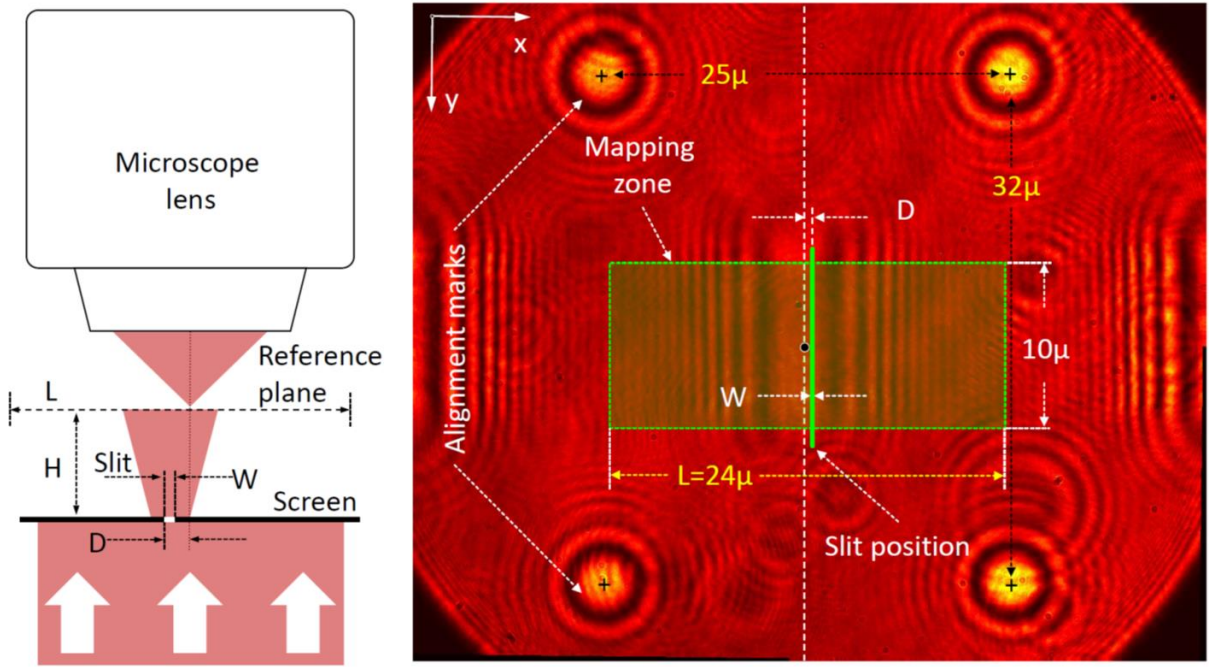

**Fig. 1.** Deeply Subwavelength Non-Contact Optical Metrology. Schematic apparatus is presented on the left (not in scale). The sample is a sub-wavelength slit in an opaque screen. A typical intensity field map recorded at distance $H = 10\lambda$ from the sample is presented on the right.

On the diffraction map, the slits are not resolved. They appear blurred and surrounded by interference fringes. The alignment marks are 1μm × 1μm squares and appear, in the diffraction pattern, as sets of concentric rings. Their positions were determined from barely resolved images at $H = 0$ using a peak finder algorithm. The sapphire substrate supports hundreds of slits of random width and position.

Single shot recording of a diffraction map was sufficient to retrieve the width $W$ and position $D$ of the slit with nanometric accuracy. They are retrieved with a deep learning artificial neural network previously trained on a set of scattering events from a number of such slits of known widths and positions. Once trained, the system is ready to measure any number of unseen slits. A dataset for training and validation has been created by fabricating a number of slits followed by recording their scattering patterns in the imaging instrument. Generating a physical set is labor-intensive, but such a set is naturally congruent with the imaging microscope ensuring high accuracy of measurements. We fabricated a set of 840 slits of random size by focused ion beam milling on a 50nm thick chromium film deposited on a sapphire substrate. In the set, widths of the slit $W$ were randomly chosen in the interval from 0.079 $\lambda$ to 0.47 $\lambda$ (50nm to 300nm). The slit position $D$ was randomized in the interval $L$ from -0.79 $\lambda$ to 0.79 $\lambda$ (-500nm to +500nm). Here, all slit widths are well below $\lambda/2$ and hence their inner structure would be considered beyond the "diffraction limit" of conventional microscopy.

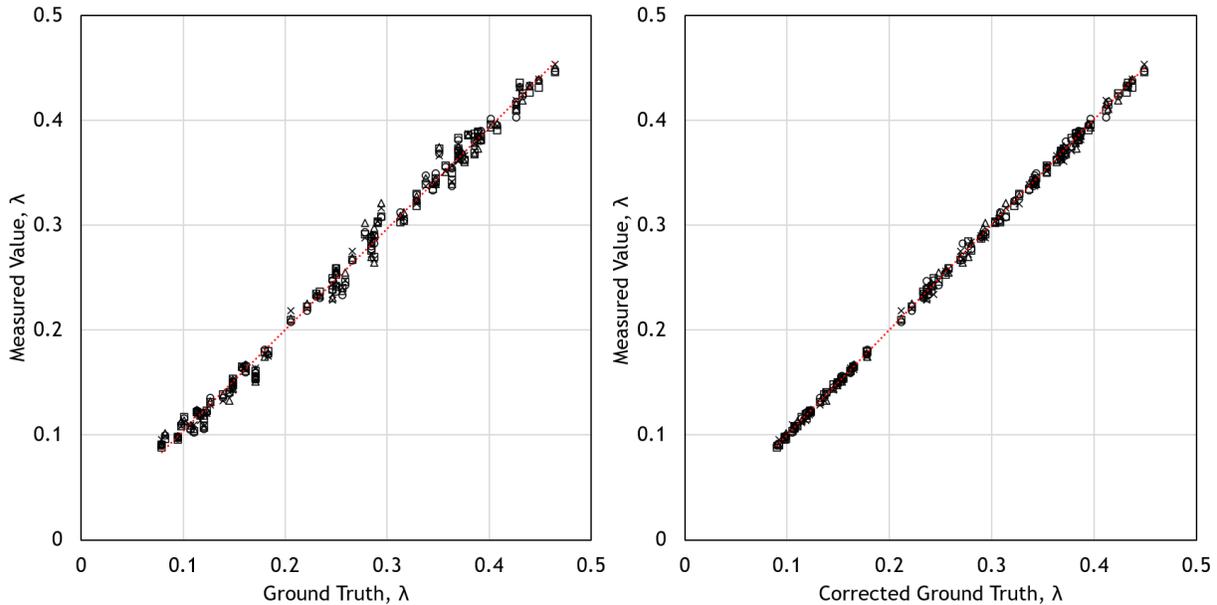

**Fig. 2.** Optical metrology of a sub-wavelength slit in the opaque screen. (Left) the measured values of width of 84 random slits against the ground truth values retrieved using the field maps at four different distances $H_k$ from the screen; (Right) The improved results of measurements based on the fabrication imperfection hypothesis. In both graphs, data for experiments taken at four different distances $H_k$ are presented: ×, $k=1$; $\triangle$, $k=2$; □, $k=3$; ○, $k=4$.

Upon completion of the training with 756 slits of *a priori* known width and position (ground truth values used for focused ion beam milling), the apparatus was ready for measuring unseen slits. To minimize errors related to the order with which the network was trained, we repeated training 100 times randomizing the training sets. The retrieved parameters of the unseen slits have been averaged over the 100 realizations of the trained networks. The standard errors of the means related to training with randomized sets *(~0.001 $\lambda$)* were negligible in comparison to other random errors in the measurement.

The results of the validation experiments on 84 randomly selected slits of unknown dimensions are presented in Fig. 2 (left). It shows the measured values of width $W$ of the slit against the ground truth values, measured in four independent experiments, at distances $H_k = 0\,\lambda,\ 2\,\lambda,\ 5\,\lambda$ and $10\,\lambda$ from the screen. Here, the retrieved values are plotted as a function of the ground truth values. The dashed red line represents perfect measurements, while deviation from the line indicates inaccuracies of the measurement. To quantify the metrology, we calculated the correlation coefficient $r_k$ between the ground truth and measured values of the slit widths, and the mean absolute error $\sigma_k$ of the measured values from the ground truth for four independent experiments taken at different distances $H_k$ (see Table 1). Here, the mean absolute error characterizes accuracy of the metrology.

**Table 1.** Accuracy of Optical Metrology

| Distance from the screen | Correlation coefficient | Accuracy, $\lambda$ | Accuracy, nm | Correlation coefficient (corrected ground truth) | Accuracy, $\lambda$ (corrected ground truth) | Accuracy, nm (corrected ground truth) |
|---|---|---|---|---|---|---|
| $H_k$ | $r_k$ | $\sigma_k$ | $\sigma_k$ | $r_k^+$ | $\sigma_k^+$ | $\sigma_k^+$ |
| $H_1 = 0\,\lambda$ | $r_1 = 0.99715$ | $\sigma_1 = 0.00770\,\lambda$ ($\lambda/130$) | 4.88nm | $r_1^+ = 0.99958$ | $\sigma_1^+ = 0.00250\,\lambda$ ($\lambda/400$) | 1.58nm |
| $H_2 = 2\,\lambda$ | $r_2 = 0.99608$ | $\sigma_2 = 0.00845\,\lambda$ ($\lambda/118$) | 5.35nm | $r_2^+ = 0.99958$ | $\sigma_2^+ = 0.00243\,\lambda$ ($\lambda/412$) | 1.54 nm |
| $H_3 = 5\,\lambda$ | $r_3 = 0.99688$ | $\sigma_3 = 0.00761\,\lambda$ ($\lambda/131$) | 4.82nm | $r_3^+ = 0.99967$ | $\sigma_3^+ = 0.00232\,\lambda$ ($\lambda/430$) | 1.47nm |
| $H_4 = 10\,\lambda$ | $r_4 = 0.99681$ | $\sigma_4 = 0.00730\,\lambda$ ($\lambda/137$) | 4.62nm | $r_4^+ = 0.99955$ | $\sigma_4^+ = 0.00249\,\lambda$ ($\lambda/401$) | 1.58nm |
| All distances | $r = 0.99673$ (Average) | $\sigma = 0.00388\,\lambda$ ($\lambda/258$) | 2.46nm | $r^+ = 0.99960$ (Average) | $\sigma^+ = 0.00122\,\lambda$ ($\lambda/821$) | 0.77nm |

We observed that retrievals of the slit width, based on mapping the diffracted fields at different distances $H_k$ from the screen, return very similar results in terms of correlation and accuracy. Measurements at all distances $H_k$ show similar error of ~ $\lambda/129$, or 4.92nm in absolute terms. As measurements at different distances are statistically independent, the accuracy of measurements statistically improves to ~ $\lambda/258$, or 2.46nm in absolute terms, when using the data collected for all four distances.

Upon completing the initial analysis, we observed that the errors of measurements taken at different distances from the screen correlate very strongly. Table 2 shows correlation coefficients $R_{kk'}$, where $k$ and $k'$ indicate experiments at different distances.

**Table 2.** Correlations of metrology results taken at different distances

| $k,k'$ | $H_1 = 0$ | $H_2 = 2\,\lambda$ | $H_3 = 5\,\lambda$ | $H_4 = 10\,\lambda$ |
|---|---|---|---|---|
| $H_1 = 0\,\lambda$ | 1 | $R_{12} = 0.90768$ ($R_{12}^+ = 0.04976$) | $R_{13} = 0.83952$ ($R_{13}^+ = -0.50666$) | $R_{14} = 0.79973$ ($R_{14}^+ = -0.58554$) |
| $H_2 = 2\,\lambda$ | | 1 | $R_{23} = 0.86700$ ($R_{23}^+ = -0.51271$) | $R_{24} = 0.83561$ ($R_{24}^+ = -0.57264$) |
| $H_3 = 5\,\lambda$ | | | 1 | $R_{34} = 0.90671$ ($R_{34}^+ = 0.13529$) |
| $H_4 = 10\,\lambda$ | | | | 1 |

We argue that such strong correlations can only be explained by accepting that our optical metrology technique gives more accurate value of the slit width than the "assumed ground truth". Indeed, if the "assumed ground truth" value is larger (smaller) than its real value, then accurate optical measurements at different distances will give obligatory correlated overestimated (underestimated) retrieved values, as illustrated in Table 2. It is not surprising, as the focused ion beam milling process used in fabrication of the slits has an instrumental tolerance caused by the finite size of the beam focus and the accuracy with which it can be positioned on the target. Therefore, we shall conclude that the average value of the measured slit width taken at different distances ("Corrected Truth") gives a more accurate value of the unknown "Real Ground Truth" than the intended value of the "assumed ground truth" used for focused ion beam milling. When assuming the average value of the measured slit width as the corrected ground truth, the correlation coefficients $r_k^+$, grow significantly, from "999" to "9999" values (see Table 1), while correlations between different measurements reduce strongly (see values $R_{k,k'}^+$, Table 2). Now the measurements show much improved average of absolute deviations $\sigma_k$ of about ~$\lambda/410$, or 1.54nm in absolute terms. Using the data collected for all four distances, the accuracy statistically improves to $\lambda/821$, or ~0.77nm in absolute terms. We therefore conclude that single shot measurements (at a given distance) have a real accuracy of ~$\lambda/410$ while the accuracy after measurements at four different distances statistically improves by the factor of 2 to $\lambda/820$.

From here, we can evaluate the average accuracy of the focused ion beam milling process in fabricating slits as $\eta = 4.75$nm. We believe that this is the first example of optical metrology beating accuracy of the focused ion beam fabrication.

To back up our claims of extraordinary level of accuracy in our optical experiments, and to help understanding the sources of errors in it, we conducted a full computer modelling experiment on the measured widths of the same 84 slits used in real validation experiments and trained the network on 756 slits as in the real experiment. In the modelling, the experimentally recorded diffraction patterns were substituted with diffraction patterns calculated using vector diffraction theory (*5*). The modelling experiment returns average accuracy for the slit width *W* measured at distances $H_i$ = 2 λ, 5 λ, 10 λ to be ~$\lambda/1304$, or 0.49nm in absolute terms. This will be compared with experimentally attainable accuracy of ~$\lambda/410$, or 1.54nm in absolute terms. The factor of 3 higher accuracy of modelling experiment in contrast with the physical experiment can be attributed to mechanical instability of the apparatus, experimental errors in aligning the field of view of the microscope to the slit and, to lesser extent, to the pixilation of image sensor.

In our real and numerical experiments, we deliberately measured slits at *a priori* unknown random position: in a practical application, correct positioning of the object in the field of view with nanometric precision is impossible. Nevertheless, the slit width is measured with high accuracy. The slit position *D* can also be measured simultaneously together with the width with deeply subwavelength accuracy.

The experimentally observed accuracy of ~$\lambda/821$ exceeds by more than two orders of magnitude the *λ/2* "diffraction limit" of conventional optical microscopes, bringing artificial intelligence enabled optical metrology to the nanometer scale accuracy that exceeds resolution of the focusing ion beam milling process with advanced tools. We therefore argue that the deep learning process involving a neural network trained on *a priori* known objects creates a powerful and accurate measurement algorithm. Remarkably, such accuracy is achieved with a small physical dataset comprising of less than a thousand slits of *a priori* known sizes. Our numerical modelling indicates that single shot sub-nanometric accuracy better than *λ/1300* will be possible, thus reaching molecular level dimensions. Moreover, using topologically structured light illumination will improve accuracy even further (*6*). However, further improvements of accuracy and precision will require larger training sets and considerable improvements in mechanical stability of the imaging apparatus.

Finally, we note that the rate at which the measurement can be performed is only limited by the camera frame rate and the neural network retrieval time. This can reach hundreds of thousands of measurements per second. The technique is simple to implement and is insensitive to where the object is placed in the field of view, the instrument does not involve moving parts and therefore it is suitable for future smart-manufacturing applications with machine tools integrated with metrology tools.

**Acknowledgments:** The authors are grateful to Nikitas Papasimakis and Bruce Ou for discussions. This work was has been supported by Singapore Ministry of Education (Grant No. MOE2016-T3-1-006), the Agency for Science, Technology and Research (A*STAR) Singapore (Grant No. SERC A1685b0005), the Engineering and Physical Sciences Research Council UK (Grants No. EP/M0091221). T.P. acknowledges the support of the China Scholarship Council (CSC No. 201804910540).